\documentclass[reqno,12pt]{amsart}
\usepackage{graphicx}
\usepackage{fullpage}

\newtheorem{thm}{Theorem}[section]

\newtheorem{lem}[thm]{Lemma}

\theoremstyle{definition}
\newtheorem{defn}[thm]{Definition}
\theoremstyle{remark}

\numberwithin{equation}{section}

\begin{document}

\title[]{Laplace Invariants for general hyperbolic systems}%
\author{Chris Athorne and Halis Yilmaz}%
\address{School of mathematics and Statistics, University of Glasgow, Glasgow, UK//
 Department of mathematics, Faculty of Education, University of Dicle, Diyarbakir, TURKEY\\}
\email{Christopher.Athorne@glasgow.ac.uk//halisyilmaz@dicle.edu.tr}%

\thanks{}%

\begin{abstract}
We consider the generalization of Laplace invariants to linear
differential systems of arbitrary rank and dimension. We discuss
completeness of certain subsets of invariants.
\end{abstract}

\maketitle
\section{Introduction}
The classical Laplace invariants \cite{L} were introduced in the
context of second order, linear hyperbolic systems of the form
\begin{equation}\label{origin}
z,_{xy}+az,_x+bz,_y+cz=0
\end{equation}
where $a$, $b$ and $c$ are given functions and $z=z(x,y)$ is an
unspecified solution of this partial differential equation.

The {\em form} of equation (\ref{origin}) is unchanged under a
general transformation $z\mapsto g(x,y)z$ where $g(x,y)$ is a
sufficiently differentiable, but otherwise arbitrary, function. In
fact the coefficients of the equation are simply mapped into new
functions,
\begin{eqnarray}
a & \mapsto & a'=a+g^{-1}g,_y\quad,\nonumber\\
b & \mapsto & b'=b+g^{-1}g,_x\quad,\nonumber\\
c & \mapsto & c'=c+g^{-1}ag,_x+g^{-1}bg,_y+g^{-1}g,_{xy}\quad\
\end{eqnarray}
and it is easily seen that the following two functions are {\em
invariant} under such a transformation:
\begin{eqnarray}
h&=&a,_x+ab-c,\\
k&=&b,_y+ab-c.
\end{eqnarray}

More than this, the pair $\{h,k\}$ is a $\it{complete}$ set of invariants
in that two equations of the form (\ref{origin}) having exactly
the same invariants, as functions of $x$ and $y$, must necessarily
be related by a gauge transformation of the sort described. The
family of equations is thus partitioned into equivalence classes
labelled by these pairs of functions.
These functions are called Laplace invariants by many researchers in integrability theory
(see e.g. \cite{Ibragimov}, \cite{jmw}, \cite{Shemyakova}, \cite{tsaousi}, \cite{tsarev2005}).

Such invariants have played an important role in recent work on
the geometrical theory of integrable systems and soliton
equations. It is not our purpose to rehearse these connections
here and we refer the interested reader to references \cite{RS,ZS}
where much of the material is reviewed. However, it is important
to point out that a valuable role is played by the {\em Laplace
map}, a differential map between equations of the form
(\ref{origin}) which acts on the equivalence classes according to
the equations of the two-dimensional Toda lattice \cite{S,W}. The
generalization of the Laplace map to higher dimension and higher
rank systems is of great importance \cite{A,ZS}. This paper should
be regarded as a prologomenon to a general theory of such
transformations.

Before proceeding let us note that the form (\ref{origin}), though
symmetric, has a degree of redundancy about it. We may choose to
transform it using a gauge transformation $z\mapsto gz$ where $g$
satisfies $g,_y=-a(x,y)g$. In this case the transformed equation
is
\begin{equation}
z,_{xy}+\int\{k-h\}dy.z,_y-hz=0,
\end{equation}
and the dependence on the equivalence class is explicit. An
equation of this reduced form,
\begin{equation}
z,_{xy}+bz,_y+cz=0,
\end{equation}
still retains a gauge covariance, namely $z\mapsto g(x)z$, the
gauge function depending upon $x$ alone and it is naturally
written as a system in $z$ and $z,_y$:
\begin{equation}
\left(\begin{array}{cc}
\partial_x & -\beta c\\
1/\beta & \partial_y
\end{array}\right)
\left(\begin{array}{c} -\beta z,_y\\z
\end{array}\right)=0
\end{equation}
where $\beta,_x=\beta b$.

Of course, we might equally consider reduced forms
\begin{equation}
z,_{xy}+az,_x+cz=0,
\end{equation}
with $y$ dependent gauge transformations, but what we cannot do in
general is reduce to the form
\begin{equation}
z,_{xy}+cz=0,
\end{equation}
as this requires that the special relationship $h=k,\,\forall x,y$
should hold.

Equally we could {\em start} with a general system form
\begin{equation}
\left(\begin{array}{cc}
\partial_x +h_{11}& h_{12}\\
h_{21}& \partial_y+h_{22}
\end{array}\right)
\left(\begin{array}{c} z_1\\z_2
\end{array}\right)=0
\end{equation}
as is done in \cite{A}. Gauge transformations preserving this form
of system are $2\times 2$ diagonal matrices acting on the two
component vector of the $z_i$. The gauge invariants are
\begin{eqnarray}
(12)&=&h_{12}h_{21},\\
{[}12{]}&=&h_{11},_y-h_{22},_x+\frac12\ln(\frac{h_{12}}{h_{21}}),_{xy}.
\end{eqnarray}
However the redundancy is also present here and we can use the
gauge transformation to kill the diagonal terms $h_{11}$ and
$h_{22}$. This leaves us with the canonical form
\begin{equation}\label{canonical}
\left(\begin{array}{cc}
\partial_x & h_{12}\\
h_{21}& \partial_y
\end{array}\right)
\left(\begin{array}{c} z_1\\z_2
\end{array}\right)=0,
\end{equation}
and residual gauge transformations
\begin{equation}
\left(\begin{array}{c} z_1\\z_2
\end{array}\right)\mapsto\left(\begin{array}{cc}
g_1(y)& 0\\
0& g_2(x)
\end{array}\right)
\left(\begin{array}{c} z_1\\z_2
\end{array}\right)
\end{equation}
with invariants
\begin{eqnarray}
(12)&=&h_{12}h_{21},\\
{[}12{]}&=&\frac12\ln(\frac{h_{12}}{h_{21}}),_{xy}.
\end{eqnarray}

It is not difficult to verify that these invariants are a $\it{complete}$
set for the canonical form (\ref{canonical}).

In what follows we shall consider $n\times n$ systems and discuss
the completeness of the sets of invariants constructed in a
similar manner to those presented in this introduction. We shall
also relate them to second order, {\em matrix} equations, i.e.
those of the type (\ref{origin}) but having $a$, $b$ and $c$ as
square matrices rather than simple functions.

We use the word {\em dimension} to denote the number of
independent variables which we shall henceforth write as
$x_1,x_2\ldots,x_n$. By {\em rank} we shall understand the number
of components in the solution vector $z$: $z_1,z_2,\ldots,z_r$.

\section{Invariants For General Hyperbolic Systems}\label{sec2}

\begin{defn}
Let ${\mathbb L}$ be an $n \times n$ matrix differential
operator
\[{\mathbb L}=\left(\begin{array}{cccc}
  \partial_1+h_{11} & h_{12}&\ldots & h_{1n}\\
   h_{21} & \partial_2+h_{22}&\ldots & h_{2n}\\
   \vdots & \vdots & \ddots & \vdots\\
   h_{n1} &h_{n2} &\ldots &\partial_{n}+h_{nn}
  \end{array}
  \right),\]
where $\partial_i$ stands for $\partial/\partial x_i$ and the $h_{ij}$ are functions of $x_1$, $x_2$,...,$x_n$. If
$g$ is a diagonal $n \times n$ matrix such that $g^{-1}$ exists, then
$H=H(h_{ij})$ is {\it invariant} under the gauge transformation
\[{\mathbb L}'=g^{-1}{\mathbb L}g,\]
so long as $H(h'_{ij})=H(h_{ij})$.
\end{defn}

\subsection{The case where rank and dimension are equal.}

In this case we deal with matrix differential operators
\begin{equation}
\mathbb L=\left(\begin{array}{cccc}
\partial_1 & 0 & \ldots & 0\\
0 & \partial_2 & & 0\\
\vdots & & \ddots & \vdots \\
0 & \ldots & & \partial_n
\end{array}\right)
+ \left(\begin{array}{cccc}
h_{11} & h_{12} & \ldots & h_{1n}\\
h_{21} & h_{22} & & h_{2n}\\
\vdots & & \ddots & \vdots \\
h_{n1} & \ldots & & h_{nn}
\end{array}\right)
\end{equation}
and gauge transformations
\begin{equation}
\mathbb L\mapsto {\mathbb L}'=g^{-1}\mathbb L g
\end{equation}
of the form
\begin{equation}
g=\left(\begin{array}{cccc}
g_1 & 0 & \ldots & 0\\
0 & g_2 & & 0\\
\vdots & & \ddots & \vdots \\
0 & \ldots & & g_n
\end{array}\right).
\end{equation}

The $h_{ij}$ and $g_i$ here are functions of all variables $x_1,
x_2,\ldots, x_n$ but we may choose the reduced (canonical) form in
which the diagonal entries $h_{11},h_{22},\ldots,h_{nn}$ are
gauged away by solving the $n$ equations:
$g_i,_i+h_{ii}g_i=0,\,i=1, 2, \ldots ,n$.

\begin{equation}\label{Canonical}
\mathbb L=\left(\begin{array}{cccc}
\partial_1 & 0 & \ldots & 0\\
0 & \partial_2 & & 0\\
\vdots & & \ddots & \vdots \\
0 & \ldots & & \partial_n
\end{array}\right)
+ \left(\begin{array}{cccc}
0 & h_{12} & \ldots & h_{1n}\\
h_{21} & 0 & & h_{2n}\\
\vdots & & \ddots & \vdots \\
h_{n1} & \ldots & & 0
\end{array}\right)
\end{equation}

The residual gauge freedom is
\begin{equation}\label{residual}
g=\left(\begin{array}{cccc}
g_1(\hat x_1) & 0 & \ldots & 0\\
0 & g_2(\hat x_2) & & 0\\
\vdots & & \ddots & \vdots \\
0 & \ldots & & g_n(\hat x_n)
\end{array}\right)
\end{equation}
where hatted variables are deleted from the list of arguments in
each $g_i$. Under such transformations
\begin{equation}
h_{ij}\mapsto g_i(\hat x_i)^{-1}g_j(\hat x_j)h_{ij}
\end{equation}
and it is easily seen that the following objects are all
invariant: Choose from the $n$ labels $\{1,2,\ldots, n\}$ a subset
of $p$ distinct ones, $\{i_1,i_2,\ldots,i_p\}$, and define the
symbol:
\begin{equation}\label{inv1}
 (i_1i_2\ldots i_p) = h_{i_1i_2}h_{i_2i_3}\ldots h_{i_pi_1}.
\end{equation}
We say the symbol $(i_1i_2\ldots i_p)$ has $\it{length}$ $p$.
Thus in the case of the symbols of lengths $2$ and $3$ we have $(ij)=h_{ij}h_{ji}$ and $(ijk)=h_{ij}h_{jk}h_{ki}$.

Because of the cyclic symmetry in these products there will be
$\frac{n!}{p(n-p)!}$ symbols of length $p$. The symbols of length
$p$ are permuted under the action of $S_n$, the symmetric group on
$n$ labels.

In addition there are $\frac12 n(n-1)$ invariants denoted by
square bracket symbols thus:
\begin{equation}\label{inv2}
 [ij]=-[ji]=\frac12\partial_i\partial_j\ln(\frac{h_{ij}}{h_{ji}}).
\end{equation}
We call the invariants (\ref{inv1}), (\ref{inv2}) $\it{simple}$.
All functions of these symbols are themselves invariant but we will now show that within the set of {\em simple} invariants
there are a $\it{complete}$ subset i.e. a set the knowledge of which is enough to determine
the operator $\mathbb L$ completely up to gauge transformations.

\begin{lem}
The functions $[ij]$ and $(i_1i_2 \ldots i_p)$ are invariants.
\end{lem}
 {\em Proof.} We consider the $n \times n$ differential
operator matrix ${\mathbb L}$
\[{\mathbb L}=\left(\begin{array}{cccc}
  \partial_1& h_{12}&\ldots & h_{1n}\\
   h_{21} & \partial_2&\ldots & h_{2n}\\
   \vdots & \vdots & \ddots & \vdots\\
   h_{n1} &h_{n2} &\ldots &\partial_{n}
  \end{array}
  \right),\]
where $h_{ij}$ are functions of $x_1$, $x_2$,..., $x_n$. We find
the invariants of ${\mathbb L}$ by using the gauge transformation,
$g^{-1}{\mathbb L}g={\mathbb L}^\prime$, where $g$ is a $n\times
n$ diagonal matrix
\[g=\left(\begin{array}{cccc}
  g_1& 0&\ldots & 0\\
   0& g_2&\ldots & 0\\
   \vdots & \vdots & \ddots & \vdots\\
   0&0&\ldots &g_n
  \end{array}
  \right).\]
Then $g^{-1}{\mathbb L}g={\mathbb L}^\prime$ gives us
\begin{eqnarray}
 0&=&(\ln g_i)_{,i},\\
 h'_{ij}&=&g^{-1}_ig_jh_{ij},~(i\neq j).
\end{eqnarray}
Now
\[\frac{1}{2}(\ln\frac{h'_{ij}}
 {h'_{ji}})_{,ij}=
 \frac{1}{2}(\ln\frac{h_{ij}}{h_{ji}})_{,ij}+(\ln\frac{g_{j}}{g_{i}})_{,ij}\]
which gives
\[\frac{1}{2}(\ln\frac{h'_{ij}}
 {h'_{ji}})_{,ij}=
 \frac{1}{2}(\ln\frac{h_{ij}}{h_{ji}})_{,ij}\]
since
\begin{eqnarray}
 g_{r,r}=0,
\end{eqnarray}
where $r=i,j$. This gives us the antisymmetric invariants
\begin{eqnarray}\label{compnxninv2}
 [ij]=\frac{1}{2}(\ln\frac{h_{ij}}{h_{ji}})_{,ij}.
\end{eqnarray}
Finally we consider the following relations
\begin{eqnarray*}
 h'_{i_1i_2}&=&g^{-1}_{i_1}g_{i_2}h_{i_1i_2}\\
 h'_{i_2i_3}&=&g^{-1}_{i_2}g_{i_3}h_{i_2i_3}\\
 h'_{i_3i_4}&=&g^{-1}_{i_3}g_{i_4}h_{i_3i_4}\\
            &.&\\
            &.&\\
h'_{i_{p-1}i_p}&=&g^{-1}_{i_{p-1}}g_{i_p}h_{i_{p-1}i_p}\\
h'_{i_pi_1}&=&g^{-1}_{i_p}g_{i_1}h_{i_pi_1}
\end{eqnarray*}
Then we obtain
\[h'_{i_1i_2}h'_{i_2i_3}h'_{i_3i_4} \ldots h'_{i_{p-1}i_p}h'_{i_pi_1}
=h_{i_1i_2}h_{i_2i_3}h_{i_3i_4} \ldots h_{i_{p-1}i_p}h_{i_pi_1}\]
to give the $p$-index invariants:
\begin{eqnarray}\label{compnxninv3}
 (i_1i_2i_3...i_p)=h_{i_1i_2}h_{i_2i_3}h_{i_3i_4} \ldots h_{i_{p-1}i_p}h_{i_pi_1},
\end{eqnarray}
where the $i_r$ are a choice of $p$ distinct integers in
$\{1,2,\ldots,n\}$.

By recalling (\ref{compnxninv2}) and (\ref{compnxninv3}) we now collect all the invariants of ${\mathbb L}$ as follows:
\begin{eqnarray*}
 \left[ij\right]&=&\frac{1}{2}(\ln\frac{h_{ij}}{h_{ji}})_{,ij}\\
  (i_1i_2i_3 \ldots i_p)&=&h_{i_1i_2}h_{i_2i_3}h_{i_3i_4} \ldots h_{i_{p-1}i_p}h_{i_pi_1}
\end{eqnarray*}

\begin{defn}
 The functions $[ij]$ and $(i_1i_2i_3 \ldots i_p)$ are called the {\em simple} invariants of ${\mathbb L}$.
\end{defn}
\begin{thm}
 {The simple invariants form a complete set for the equivalence class of ${\mathbb L}$ under gauge transformations,
  where ${\mathbb L}$ is defined by (\ref{Canonical})}.
\end{thm}

{\em Proof.} The proof depends on showing that one can construct a
suitable gauge matrix $g$. In other words we need to show that
\begin{eqnarray*}
 {\mathbb L}^{\prime}=g^{-1}{\mathbb L}g\Longleftrightarrow
 \left\{\begin{array}{ccc}
   \left[ij\right]'&=&\left[ij\right]\\
   (i_1i_2i_3 \ldots i_p)'&=&(i_1i_2i_3 \ldots i_p)
 \end{array}\right\}
\end{eqnarray*}
where $\{i_1,i_2,i_3, \ldots ,i_p\} \subset \{1,2, \ldots ,n\}$.

We already know that `$\Rightarrow$'  is true. We only need to
prove the `$\Leftarrow$' part. Assume the RHS is true i.e.
\begin{eqnarray*}
   \left[ij\right]'=\left[ij\right]\\
   (i_1i_2i_3 \ldots i_p)'&=&(i_1i_2i_3 \ldots i_p),
\end{eqnarray*}
for all subsets $\left\{i_1,i_2, \ldots ,i_p\right\}\subseteq \{1,2,\ldots , n\}$.
Let us choose an $n
\times n$ diagonal matrix $f$ such that
\[f=\left(\begin{array}{cccc}
  f_1& 0& \ldots & 0\\
   0& f_2& \ldots & 0\\
   \vdots & \vdots & \ddots & \vdots\\
   0&0&\ldots &f_n
  \end{array}
  \right),\]
where
\begin{eqnarray*}
 f_1&=&h_{12}h_{23}h_{34} \ldots h_{n-1n},\\
 f_2&=&h'_{12}h_{23}h_{34} \ldots h_{n-1n},\\
 f_3&=&h'_{12}h'_{23}h_{34} \ldots h_{n-1n},\\
    &.&\\
    &.&\\
 f_{n-1}&=&h'_{12}h'_{23}h'_{34} \ldots h'_{n-2n-1}h_{n-1n},\\
 f_n&=&h'_{12}h'_{23}h'_{34} \ldots h'_{n-2n-1}h'_{n-1n} .
\end{eqnarray*}
Then we obtain
\[f^{-1}{\mathbb L}f=\left(\begin{array}{cccc}
  \partial_1+\tilde{h}_{11}& \tilde{h}_{12}&\ldots & \tilde{h}_{1n}\\
   \tilde{h}_{21} & \partial_2+\tilde{h}_{22}&\ldots & \tilde{h}_{2n}\\
   \vdots & \vdots & \ddots & \vdots\\
   \tilde{h}_{n1} &\tilde{h}_{n2} &\ldots &\partial_{n}+\tilde{h}_{nn}
  \end{array}
  \right),\]
where
\begin{eqnarray*}
 \tilde{h}_{ij}&=&f^{-1}_if_jh_{ij}\quad (i\neq j),\\
 \tilde{h}_{ii}&=&(\ln f_i)_{,i} \quad (i=1,2, \ldots ,n).
\end{eqnarray*}
Thus we need to show
\begin{eqnarray}\label{tildeh}
 \tilde{h}_{ij}=h'_{ij}\text{~~}(i\neq j).
\end{eqnarray}
We easily prove (\ref{tildeh}) as follows:
\begin{eqnarray*}
 \tilde{h}_{ij}&=&f^{-1}_if_jh_{ij}\text{~~}(i\neq j)\\
\end{eqnarray*}
Let $i<j$. Then
\begin{eqnarray*}
 f_i&=&h'_{12}h'_{23} \ldots h'_{i-1i}h_{ii+1} \ldots h_{j-1j}h_{jj+1} \ldots h_{n-1n}\\
 f_j&=&h'_{12}h'_{23} \ldots h'_{i-1i}h'_{ii+1} \ldots h'_{j-1j}h_{jj+1} \ldots h_{n-1n}
\end{eqnarray*}
Thus
\begin{eqnarray*}
 \tilde{h}_{ij}=\frac{f_j}
 {f_i}h_{ij}&=&\frac{h'_{ii+1}h'_{i+1i+2} \ldots h'_{j-1j}}{h_{ii+1}h_{i+1i+2} \ldots h_{j-1j}}h_{ij}\\\\
            &=&\frac{h'_{ii+1}h'_{i+1i+2} \ldots h'_{j-1j}h'_{ji}}{h_{ii+1}h_{i+1i+2} \ldots h_{j-1j}h_{ji}}.\frac{h_{ji}}{h'_{ji}}.h_{ij}\\\\
            &=&\frac{(ii+1i+2 \ldots j)'}{(ii+1i+2 \ldots j)}.\frac{(ij)}{h'_{ji}}.\frac{h'_{ij}}{h'_{ij}}\\\\
            &=&\frac{(ij)}{(ij)'}.h'_{ij}\\\\
            &=&h'_{ij}
\end{eqnarray*}
since $(ii+1i+2 \ldots j)'=(ii+1i+2 \ldots j)$ and $(ij)'=(ij)$.

Similarly
\begin{eqnarray*}
 \tilde{h}_{ji}=\frac{f_i}
 {f_j}h_{ji}&=&\frac{h_{ii+1}h_{i+1i+2} \ldots h_{j-1j}}{h'_{ii+1}h'_{i+1i+2} \ldots h'_{j-1j}}h_{ji}\\\\
 &=&\frac{(ii+1i+2 \ldots j)}{(ii+1i+2 \ldots j)'}h'_{ji}\\\\
 &=&h'_{ji}.
\end{eqnarray*}

Hence for $i\neq j$ we obtain
\begin{eqnarray*}
 \tilde{h}_{ij}=h'_{ij}.
\end{eqnarray*}

So we have
\[f^{-1}{\mathbb L}f=\left(\begin{array}{cccc}
  \partial_1+\tilde{h}_{11}& h'_{12}&\ldots & h'_{1n}\\
   h'_{21} & \partial_2+\tilde{h}_{22}&\ldots & h'_{2n}\\
   \vdots & \vdots & \ddots & \vdots\\
   h'_{n1} &h'_{n2} &\ldots &\partial_{n}+\tilde{h}_{nn}
  \end{array}
  \right)\]
where
\begin{eqnarray}\label{tildehii}
 \tilde{h}_{ii}&=&(\ln f_i)_{,i}.
\end{eqnarray}
We now need to seek a single function $\theta$ so that
\begin{eqnarray*}
 \theta^{-1}(f^{-1}{\mathbb L}f)\theta={\mathbb L}'.
\end{eqnarray*}
This requires that $\theta$ satisfy the following equations:
\begin{eqnarray*}
 \theta^{-1} \theta_{,i}+\tilde{h}_{ii}=0
\end{eqnarray*}
i.e.
\begin{eqnarray*}
 \theta_{,i}=-\tilde{h}_{ii}\theta.
\end{eqnarray*}
The above equations are consistent $\Longleftrightarrow
(\theta_{,i})_{,j}=(\theta_{,j})_{,i}$, which gives
\begin{eqnarray}\label{conshiijj}
 \tilde{h}_{ii,j}=\tilde{h}_{jj,i}.
\end{eqnarray}
Recalling (\ref{tildehii}) we write
\begin{eqnarray*}
 \tilde{h}_{ii}=\left(\ln f_i\right)_{,i},\\
 \tilde{h}_{jj}=\left(\ln f_j\right)_{,j},
\end{eqnarray*}
and if we substitute these into the equation (\ref{conshiijj}) we
obtain
\begin{eqnarray*}
 \left[ij\right]'=\left[ij\right]
\end{eqnarray*}
since
\[\left(\frac{f_i}{f_j}\right)^2=\frac{h_{ij}}{h_{ji}}\frac{h'_{ji}}{h'_{ij}},\]
where
\begin{eqnarray*}
\frac{f_i}{f_j}=\frac{h_{ij}}{h'_{ij}}=\frac{h'_{ji}}{h_{ji}}.
\end{eqnarray*}
So the equality of invariants guarantees that the Frobenius
integrability condition is satisfied: there exists a function
$\theta$ such that $\theta^{-1}(f^{-1}{\mathbb L}f)\theta={\mathbb
L}'$ i.e.
\[g^{-1}{\mathbb L}g={\mathbb L}',\]
where $g=\theta f$. Hence the given invariants of ${\mathbb L}$
are a complete set.

\vspace{.2in} It should be noted that the simple invariants are
not algebraically independent. For instance,
\begin{equation}
 (ijk)(ikj)=(ij)(jk)(ki)
\end{equation}
so that there must be a smallest set of simple invariants which is
still complete. A {\em minimal complete} set is given in the
following result:

\begin{thm}\label{thrmin.inv}
The simple invariants $(1i)$, $[ij]$ and $(1ij)$ form a minimal complete set.
\end{thm}
First we prove some lemmas.
\begin{lem}\label{lm1}
 We consider a simple invariant of length $m$
\begin{eqnarray}
(i_1i_2i_3 \ldots i_{m-1}i_m)&=&h_{i_1i_2}h_{i_2i_3}h_{i_3i_4} \ldots h_{i_{m-1}i_m}h_{i_mi_1}.
\end{eqnarray}
 Let $m$ be a positive integer such that $m\geq4$. Then
\end{lem}
\begin{eqnarray}
 (i_1i_2i_3 \ldots i_{m-1}i_m)=\frac{(i_1i_2i_3 \ldots i_{m-1})(i_1i_{m-1}i_m)}{(i_1i_{m-1})}
\end{eqnarray}
{\em Proof.}
\begin{eqnarray*}
 RHS&=&\frac{(i_1i_2i_3 \ldots i_{m-1})(i_1i_{m-1}i_m)}{(i_1i_{m-1})}\\
    &=&\frac{h_{i_1i_2}h_{i_2i_3} \ldots h_{i_{m-2}i_{m-1}}h_{i_{m-1}i_1}.h_{i_1i_{m-1}}h_{i_{m-1}i_m}h_{i_mi_1}}{h_{i_1i_{m-1}}h_{i_{m-1}i_1}}\\
    &=&h_{i_1i_2}h_{i_2i_3} \ldots h_{i_{m-2}i_{m-1}}h_{i_{m-1}i_m}h_{i_mi_1}\\
    &=&(i_1i_2i_3 \ldots i_{m-1}i_m)=LHS
\end{eqnarray*}
Hence we can replace simple invariants of length $m\geq4$
with invariants of length $m-1$ up to multiples of invariants of lengths $2$ and $3$.
\begin{lem}\label{lm2}
 Let $i,j,k$ be three positive integers such that $i$$\neq$$j$$\neq$$k$ . Then
\end{lem}
\begin{eqnarray}
 (ij)&=&\frac{(1ij)(1ji)}{(1i)(1j)}\\
 (ijk)&=&\frac{(1ij)(1jk)(1ki)}{(1i)(1j)(1k)}
\end{eqnarray}
{\em Proof.}

\begin{eqnarray*}
   \frac{(1ij)(1ji)}{(1i)(1j)}
    &=&\frac{h_{1i}h_{ij}h_{j1}.h_{1j}h_{ji}h_{i1}}{h_{1i}h_{i1}.h_{1j}h_{j1}}\\
    &=&h_{ij}h_{ji}=(ij)=LHS
\end{eqnarray*}
Similarly
\begin{eqnarray*}
 \frac{(1ij)(1jk)(1ki)}{(1i)(1j)(1k)}
    &=&\frac{h_{1i}h_{ij}h_{j1}.h_{1j}h_{jk}h_{k1}.h_{1k}h_{ki}h_{i1}}{h_{1i}h_{i1}.h_{1j}h_{j1}.h_{1k}h_{k1}}\\
    &=&h_{ij}j_{jk}h_{ki}\\
    &=&(ijk)=LHS
\end{eqnarray*}

\begin{lem}\label{lm3}
  The invariants $(1ij)$ are irreducible (i.e. they cannot be written purely in terms of invariants with length 2)
\end{lem}
{\em Proof.} We will prove this by contradiction. So assume $(1ij)$ is reducible. Thus $(1ij)$ can be expressed in terms of the invariants
$(1i)$, $(1j)$ and $(ij)$. So let
\begin{eqnarray}\label{eqn1ijF}
 (1ij)=F[(1i),(1j),(ij)]
\end{eqnarray}
If we differentiate the equation (\ref{eqn1ijF}) with respect to $h_{i1},h_{1j}$ and $h_{ji}$ respectively
we obtain the following partial differential equations:
\begin{eqnarray*}
 0=\frac{\partial(1ij)}{\partial h_{i1}}&=&\frac{\partial F}{\partial(1i)}.h_{1i}\\
 0=\frac{\partial(1ij)}{\partial h_{1j}}&=&\frac{\partial F}{\partial(1j)}.h_{j1}\\
 0=\frac{\partial(1ij)}{\partial h_{ji}}&=&\frac{\partial F}{\partial(ij)}.h_{ij}
\end{eqnarray*}
since $(1ij)=h_{1i}h_{ij}h_{j1}$ is independent of $h_{i1}$, $h_{1j}$ and $h_{ji}$.

Thus, we find
\begin{eqnarray*}
 \frac{\partial F}{\partial(1i)}=0\\
 \frac{\partial F}{\partial(1j)}=0\\
 \frac{\partial F}{\partial(ij)}=0
\end{eqnarray*}
since $h_{1i}\neq0, h_{j1}\neq0$ and $h_{ij}\neq0$.

This shows that $(1ij)= constant$. This is a contradiction. Therefore the invariant $(1ij)$ is irreducible.

{\em Proof of Theorem \ref{thrmin.inv}.}
We have considered the following simple invariants of length $m$:
\begin{eqnarray*}
(i_1i_2i_3...i_{m-1}i_m)&=&h_{i_1i_2}h_{i_2i_3}h_{i_3i_4}...h_{i_{m-1}i_m}h_{i_mi_1}
\end{eqnarray*}
First we have shown (Lemma \ref{lm1}) that these invariants can be reduced up to length $3$ and then
we have shown (Lemma \ref{lm2}) that the invariant $(ij)$ can be written in terms of the simple invariants $(1i)$ and $(1ij)$
and we have also proved that the simple invariant $(ijk)$ can be expressed in terms of the invariants $(1i)$ and $(1ij)$.
Finally, we have proved (Lemma \ref{lm3}) that the invariant $(1ij)$ is not reducible, in other words,
it can not be reduced to the invariant of length $2$.

Hence the proof of the theorem is complete and the result
follows: Any invariant of length $m$ can be written in terms of
the minimal invariants $(1i)$ and $(1ij)$ where these minimal invariants together with $[ij]$ form a complete set.


\section{Matrix Covariants For General Hyperbolic Systems}\label{sec3}

\subsection{Matrix Covariants}

Let us consider the system
\begin{eqnarray}\label{hyp.exn.cov}
 Lz=(\partial_x\partial_y+a\partial_x+b\partial_y+c)z=0
\end{eqnarray}
where $a$, $b$ and $c$ are $m \times m$ square matrices. This case is considered in \cite{Kon}. The gauge transformation on the differential operator $L$
is $L'=g^{-1}Lg$, where $g$ is a $m \times m$ diagonal matrix which gives
\begin{equation}\label{konp}
 \begin{split}
 h=a_{,x}+ba-c\\
 k=b_{,y}+ab-c
 \end{split}
\end{equation}
where $h$ and $k$ are gauge {\em covariants}\text{~} for the system (\ref{hyp.exn.cov}):
$h'=g^{-1}hg$, $k'=g^{-1}kg$. These covariants are sometimes called invariants in the literature \cite{Kon}.

\subsection{Matrix Covariants for ${\mathbb L}$}

Let us consider ${\mathbb L}$ as a $(m_1+m_2)\times(m_2+m_1)$ differential matrix operator such that
 \[{\mathbb L}=\left(\begin{array}{cc}
   \partial_1+h_{11} & h_{12}\\
   h_{21} & \partial_2+h_{22}
  \end{array}
  \right)\]
 where $h_{11}\in M_{m_1m_1}$, $h_{12}\in M_{m_1m_2}$, $h_{21}\in M_{m_2m_1}$, $h_{22}\in M_{m_2m_2}$ and
 $M_{m_im_j}$ is the set of $m_i \times m_j$ matrices.

 Strictly speaking we should write $I_{m_1}\partial_1$ and $I_{m_2}\partial_2$ for the differential operator
 entries where $I_{m_1}$, $I_{m_2}$ are the unit matrices of dimensions $m_1$ and $m_2$. This should be understood
 in what follows.

The `gauge' transformation $g$ on ${\mathbb L}$ is ${\mathbb L}'=g^{-1}{\mathbb L}g$ for
\[g=\left(\begin{array}{cc}
  g_1 & 0\\
   0 & g_2
  \end{array}
  \right)\]
where $g_1\in M_{m_1m_1}$ and $g_2\in M_{m_2m_2}$ are both invertible square matrix functions of
$x_1,x_2$. Under this action, ${\mathbb L}'=g^{-1}{\mathbb L}g$, we have
\begin{eqnarray*}
 h_{11}'&=&g_1^{-1}h_{11}g_1+g_{1}^{-1}g_{1,1},\\
 h_{12}'&=&g_{1}^{-1}h_{12}g_2,\\
 h_{21}'&=&g_{2}^{-1}h_{21}g_1,\\
 h_{22}'&=&g_{2}^{-1}h_{22}g_2+g_{2}^{-1}g_{2,2}.
\end{eqnarray*}
\subsection{Definitions}
We call an object $H$ of {\em type} $G_i \times G_j$ if $H'=g_i^{-1}Hg_j$. Therefore $h_{12}$ is
of type $G_1 \times G_2$ and $h_{21}$ is of type $G_2 \times G_1$. Covariants are of type $G_i \times G_i$.
In other words $H$ is a $\it{covariant}$ if $H'=g_i^{-1}Hg_i$. Invariants are given by the traces of covariants.
The operators $\partial_1+h_{11}$ and $\partial_2+h_{22}$ are of types $G_1 \times G_1$ and $G_2 \times G_2$ respectively:
\begin{eqnarray*}
 \partial_1+h_{11}'&=&g_1^{-1}(\partial_1+h_{11})g_1,\\
 \partial_2+h_{22}'&=&g_2^{-1}(\partial_2+h_{22})g_2.
\end{eqnarray*}
But they are $\it{differential\text{~}operator\text{~} covariants}$. We seek $\it{matrix\text{~} covariants}$. The simplest matrix
covariants are $h_{12}h_{21}$ of type $G_1 \times G_1$ and $h_{21}h_{12}$ of type $G_2 \times G_2$  since,
\begin{eqnarray*}
 h_{12}'h_{21}'&=&g_1^{-1}(h_{12}h_{21})g_1,\\
 h_{21}'h_{12}'&=&g_2^{-1}(h_{21}h_{12})g_2.
\end{eqnarray*}
Let us call ${\bf(12)}=h_{12}h_{21}$ and ${\bf(21)}=h_{21}h_{12}$, where ${\bf(12)} \in M_{m_1m_1}$
and ${\bf(21)} \in M_{m_2m_2}$. Note that we use similar notation to before but now ${\bf(12)}\neq{\bf(21)}$.
Our aim is now to form higher matrix covariants. For simplicity we call $\partial_1+h_{11}=D_1$ and
$\partial_2+h_{22}=D_2$. The operators $D_1$ and $D_2$ are of type $G_1\times G_1$ and $G_2\times G_2$ respectively.
Therefore one easily see that
\begin{eqnarray*}
h_{21}D_1 \text{~and~} D_2h_{21} \text{~are of type~} G_2\times G_1,\\
h_{12}D_2 \text{~and~} D_1h_{12} \text{~are of type~} G_1\times G_2.
\end{eqnarray*}
Hence
\begin{eqnarray*}
 c_{11}=h_{12}D_2h_{21}D_1-D_1h_{12}D_2h_{21} \text{~is type of~}G_1\times G_1,\\
 c_{22}=h_{21}D_1h_{12}D_2-D_2h_{21}D_1h_{12} \text{~is type of~}G_2\times G_2.
\end{eqnarray*}
But these are still not matrix covariants, since they have leading differential operator terms
\begin{eqnarray*}
 c_{11}=-[D_1,{\bf(12)}]\partial_2+ \ldots\\
 c_{22}=-[D_2,{\bf(21)}] \partial_1+\ldots
\end{eqnarray*}

We would like to subtract off multiples of $\partial_2+h_{22}$ from $c_{11}$
and $\partial_1+h_{11}$ from $c_{22}$ to remove the differential operators but each operator is of the wrong type.
To circumvent this we turn $c_{11}$, $c_{22}$ into respectively $G_2\times G_2$ and $G_1\times G_1$ of type covariants by:
\begin{eqnarray}
 h_{21}c_{11}h_{12}&=&-h_{21}\left[D_1,{\bf(12)}\right]\partial_2h_{12}+\ldots\text{matrix}\nonumber\\
 &=&-h_{21}[D_1,{\bf(12)}]h_{12}(\partial_2+h_{22})+\ldots\text{matrix},\label{hc11}\\
 h_{12}c_{22}h_{21}&=&-h_{12}[D_2,{\bf(21)}]h_{21}(\partial_1+h_{11})+\ldots\text{matrix}.\label{hc22}
\end{eqnarray}
Since each part in expression (\ref{hc11}) is now of type $G_2\times G_2$ and each part in (\ref{hc22}) of type
$G_1\times G_1$, we must have matrix covariants:
\begin{eqnarray*}
 {\bf[12]}=h_{12}c_{22}h_{21}+h_{12}[D_2,{\bf(21)}]h_{21}(\partial_1+h_{11}),\\
 {\bf[21]}=h_{21}c_{11}h_{12}+h_{21}[D_1,{\bf(12)}]h_{12}(\partial_2+h_{22}).
\end{eqnarray*}
Simplifying these give
\begin{eqnarray}
 {\bf[12]}={\bf(12)}[D_1,h_{12}D_2h_{21}]-h_{12}D_2h_{21}[D_1,{\bf(12)}]\text{~of~type~}G_1\times G_1\\
 {\bf[21]}={\bf(21)}[D_2,h_{21}D_1h_{12}]-h_{21}D_1h_{12}[D_2,{\bf(21)}]\text{~of~type~}G_2\times G_2
\end{eqnarray}
as matrix covariants where ${\bf[12]}\in M_{m_1m_1}$ and ${\bf[21]} \in M_{m_2m_2}$.

{\bf The case $m_1=m_2=1$:}

We find a reduction of $\bf[12]$ and $\bf[21]$ in the case $m_1=m_2=1$. So in this case ${\bf(12)}=h_{12}h_{21}$
and ${\bf(21)}=h_{21}h_{12}$ are just equal functions and ${\bf(21)}={\bf(12)}=(12)$, the earlier invariant.
By substituting $D_1=\partial_1+h_{11}$ and $D_2=\partial_2+h_{22}$ in the covariants $\bf[12]$, $\bf[21]$
and then by doing some differential and algebraic calculations we obtain the function covariants as follows:
\begin{equation}\label{f.covts1}
 \begin{split}
 {\bf[12]}=-\frac{1}{4}(12)^2_{,12}-(12)^2[12],\\
 {\bf[21]}=-\frac{1}{4}(12)^2_{,12}+(12)^2[12].
 \end{split}
\end{equation}
One easily sees that
\begin{equation}\label{f.covts2}
 \begin{split}
  {\bf[12]}+{\bf[21]}=-\frac{1}{2}(12)_{,12}^2,\\
  {\bf[21]}-{\bf[12]}=2(12)^2[12] \hspace{0.1cm}
 \end{split}
\end{equation}
Thus relating the expressions from the new covariants to the old invariants in this case $(m_1=m_2=1)$.

\subsection{The case where rank exceeds dimension.}

It is clear that in the case where the rank $r$ is larger than the dimension $n$
we may attempt to repeat the arguments of section \ref{sec2} under the weaker hypothesis
that the $h_{ij}$ and $g_i$ are matrices and no longer (commuting) functions.
The canonical form (\ref{Canonical}) still suffices where now the
$h_{ij}$ are rectangular matrices of type $m_i\times m_j$, where an $m_i\times m_i$ unit matrix
is taken to stand (but omitted) before each operator, $\partial_i$, and where $m_1+m_2+\ldots+m_n=r$.

{\bf{The case} $n=2$:}

In this case we consider a differential matrix operator ${\mathbb L}$ such that
\begin{eqnarray}
 {\mathbb L}=\left(\begin{array}{cc}
  \partial_1 & h_{12}\\
   h_{21} & \partial_2
  \end{array}
  \right)
\end{eqnarray}
where $h_{12}\in M_{m_1m_2}$ and $h_{21}\in M_{m_2m_1}$
are matrix functions of $x_1$ and $x_2$.
We have assumed a gauge transformation to this form as before.

The gauge transformation
\begin{equation} \mathbb L\mapsto
{\mathbb L}'=g^{-1}\mathbb L g,
\end{equation}
where
\begin{eqnarray}
 g=\left(\begin{array}{cc}
  g_1(x_2)& 0\\
   0 & g_2(x_1)
  \end{array}
  \right),
\end{eqnarray}
gives us
\begin{eqnarray}
 h_{12}\mapsto h'_{12}=g_{1}^{-1}h_{12}g_2 \label{matrixh12}\\
 h_{21}\mapsto h'_{21}=g_{2}^{-1}h_{21}g_1\label{matrixh21}
\end{eqnarray}
where $g_1$ and $g_2$ are invertible square matrices such that
$g_1\in M_{m_1m_1}$ and $g_2\in M_{m_2m_2}$.\\
So the relations (\ref{matrixh12}) and (\ref{matrixh21}) give us
\begin{eqnarray*}
 h'_{12}h'_{21}=g_{1}^{-1}\left(h_{12}h_{21}\right)g_1\\
 h'_{21}h'_{12}=g_{2}^{-1}\left(h_{21}h_{12}\right)g_2
\end{eqnarray*}
Thus we have
\begin{eqnarray}
 {\bf(12)}'=g_1^{-1}{\bf(12)} g_1\\
 {\bf(21)}'=g_2^{-1}{\bf(21)} g_2
\end{eqnarray}
where ${\bf(12)} \in M_{m_1m_1}$ and ${\bf(21)} \in M_{m_2m_2}$ are matrix covariants such that
\begin{eqnarray}
 {\bf(12)}=h_{12}h_{21}\\
 {\bf(21)}=h_{21}h_{12}
\end{eqnarray}
By doing some algebraic calculations over (\ref{matrixh12}) and
(\ref{matrixh21}) we obtain
\begin{eqnarray*}
 {\bf(12)}'\left(h'_{12,2}h'_{21}\right)_{,1}+h'_{12}h'_{21,2}{\bf(12)}'_{,1}
 =g_1^{-1}\left({\bf(12)}\left(h_{12,2}h_{21}\right)_{,1}+h_{12}h_{21,2}{\bf(12)}_{,1}\right)g_1\\
 {\bf(21)}'\left(h'_{21,1}h'_{12}\right)_{,2}+h'_{21}h'_{12,1}{\bf(21)}'_{,2}
 =g_2^{-1}\left({\bf(21)}\left(h_{21,1}h_{12}\right)_{,2}+h_{21}h_{12,1}{\bf(21)}_{,2}\right)g_2
\end{eqnarray*}
Therefore we have
\begin{eqnarray}
   {\bf[12]}'=g_1^{-1}{\bf[12]}g_1\\
   {\bf[21]}'=g_2^{-1}{\bf[21]}g_2
\end{eqnarray}
where we define matrix covariants ${\bf[12]} \in M_{m_1m_1}$ and ${\bf[21]} \in M_{m_2m_2}$ as follows
\begin{eqnarray}
 {\bf[12]}={\bf(12)}\left(h_{12,2}h_{21}\right)_{,1}+h_{12}h_{21,2}{\bf(12)}_{,1}\\
 {\bf[21]}={\bf(21)}\left(h_{21,1}h_{12}\right)_{,2}+h_{21}h_{12,1}{\bf(21)}_{,2}
\end{eqnarray}
Before we move to the case $n=3$, we compare our covariants ${\bf(12)}, {\bf(21)}, {\bf[12]}, {\bf[21}]$
with Konopelchenko's covariants (\ref{konp}): $h=a_{,x}+ba-c$, $k=b_{,y}+ab-c$, where $h$ and $k$ are covariants for the hyperbolic system $z_{xy}+az_x+bz_y+cz=0$. This corresponds to $m_1=m_2$ in the current context.
As we already know this system can be written in a differential operator form as
$Lz=\left(\partial_x\partial_y+a\partial_x+b\partial_y+c\right)z=0$,
where the differential operator $L=\partial_x\partial_y+a\partial_x+b\partial_y+c$ can be written as
\begin{eqnarray*}
 L&=&(\partial_x+b)(\partial_y+a)-h\\
  &=&(\partial_y+a)(\partial_x+b)-k.
\end{eqnarray*}
Therefore, we can rewrite the above system $Lz=0$ as
\begin{eqnarray}
 Au=0 \label{sysA}\\
 Bv=0 \label{sysB}
\end{eqnarray}
where
\begin{eqnarray*}
 A=\left(\begin{array}{cc}
  \partial_x+b & -h\\
   -I & \partial_y+a
  \end{array}
  \right),
  u=\left(\begin{array}{c}
   z_1\\
   z
 \end{array}\right);~~~
  B=\left(\begin{array}{cc}
  \partial_x+b & -I\\
   -k & \partial_y+a
  \end{array}
  \right),
  v=\left(\begin{array}{c}
   z\\
   z_2
 \end{array}\right).
\end{eqnarray*}
For the system (\ref{sysA}), we obtain covariant relations:
\begin{equation}\label{relation1}
 \begin{split}
 {\bf(12)}={\bf(21)}=h \\
 {\bf[12]}=hh_{xy} \hspace{0.7cm}\\
 {\bf[21]}=h_xh_y \hspace{0.7cm}
 \end{split}
\end{equation}
where $m_1=m_2=m$ and $\partial_1=\partial_x,~ \partial_2=\partial_y$.\\
We can easily see that
\begin{eqnarray}
 {\bf[12]}+{\bf[21]}=\frac{1}{2}\left(hh_y\right)_{x}
\end{eqnarray}
Thus, we have
\begin{eqnarray}\label{trh}
 Tr{\bf[12]}+Tr{\bf[21]}=\frac{1}{2}\left[Tr {\bf(12)(21)}\right]_{xy}
\end{eqnarray}
Similarly, for the system (\ref{sysB}), we have the following relations:
\begin{equation}\label{relation2}
 \begin{split}
 {\bf(12)}={\bf(21)}=k \\
 {\bf[12]}=k_yk_x \hspace{0.7cm}\\
 {\bf[21]}=kk_{xy} \hspace{0.7cm}
 \end{split}
\end{equation}
These relations give us
\begin{eqnarray}
 {\bf[12]}+{\bf[21]}=\frac{1}{2}\left(kk_x\right)_{y}
\end{eqnarray}
Once again, we have
\begin{eqnarray}
 Tr{\bf[12]}+Tr{\bf[21]}=\frac{1}{2}\left[Tr {\bf(12)(21)}\right]_{xy}
\end{eqnarray}

{\bf{The case} $n=3$:}

Here we consider a differential matrix operator ${\mathbb L}$ such that
\begin{eqnarray}
 {\mathbb L}=\left(\begin{array}{ccc}
  \partial_1 & h_{12} & h_{13}\\
   h_{21} & \partial_2 & h_{23}\\
   h_{31} & h_{32} & \partial_3
  \end{array}
  \right)
\end{eqnarray}
where $h_{ij}\in M_{m_im_j}$ $(i,j=1,2,3)$ are functions of $x_1$, $x_2$ and $x_3$.

Applying the gauge transformation
\begin{equation} \mathbb L\mapsto
{\mathbb L}'=g^{-1}\mathbb L g,
\end{equation}
where
\begin{eqnarray}
 g=\left(\begin{array}{ccc}
  g_1(x_2,x_3)& 0 & 0\\
  0 & g_2(x_1,x_3) & 0\\
  0 & 0 & g_3(x_1,x_2)
  \end{array}
  \right),
\end{eqnarray}
gives us
\begin{eqnarray}
  h'_{12}=g_{1}^{-1}h_{12}g_2,~~~~~ h'_{13}=g_{1}^{-1}h_{13}g_3\label{matrixh1}\\
  h'_{21}=g_{2}^{-1}h_{21}g_1,~~~~~ h'_{23}=g_{2}^{-1}h_{23}g_3\label{matrixh2}\\
  h'_{31}=g_{3}^{-1}h_{31}g_1,~~~~~ h'_{32}=g_{3}^{-1}h_{32}g_2\label{matrixh3}
\end{eqnarray}
where $g_1$, $g_2$ and $g_3$ are invertible square matrices such that
$g_1\in M_{m_1m_1}$, $g_2\in M_{m_2m_2}$ and $g_3\in M_{m_3m_3}$.

By doing some algebraic calculation over the above relations (\ref{matrixh1}) -- (\ref{matrixh3}),
we obtain the following matrix covariants:
\begin{eqnarray}
 &&{\bf(12)}=h_{12}h_{21}, ~~~~~ {\bf(13)}=h_{13}h_{31}\\
 &&{\bf(23)}=h_{23}h_{32}, ~~~~~ {\bf(21)}=h_{21}h_{12}\\
 &&{\bf(31)}=h_{31}h_{13}, ~~~~~ {\bf(32)}=h_{32}h_{23}\\
 &&{\bf(123)}=h_{12}h_{23}h_{31}, ~~~~~ {\bf(132)}=h_{13}h_{32}h_{21}\\
 &&{\bf(231)}=h_{23}h_{31}h_{12}, ~~~~~ {\bf(213)}=h_{21}h_{13}h_{32}\\
 &&{\bf(312)}=h_{31}h_{12}h_{23}, ~~~~~ {\bf(321)}=h_{32}h_{21}h_{13}\\
 &&{\bf[12]}={\bf(12)}\left(h_{12,2}h_{21}\right)_{,1}+h_{12}h_{21,2}{\bf(12)}_{,1}\\
 &&{\bf[13]}={\bf(13)}\left(h_{13,3}h_{31}\right)_{,1}+h_{13}h_{31,3}{\bf(13)}_{,1}\\
 &&{\bf[21]}={\bf(21)}\left(h_{21,1}h_{12}\right)_{,2}+h_{21}h_{12,1}{\bf(21)}_{,2}\\
 &&{\bf[23]}={\bf(23)}\left(h_{23,3}h_{32}\right)_{,2}+h_{23}h_{32,3}{\bf(23)}_{,2}\\
 &&{\bf[31]}={\bf(31)}\left(h_{31,1}h_{13}\right)_{,3}+h_{31}h_{13,1}{\bf(31)}_{,3}\\
 &&{\bf[32]}={\bf(32)}\left(h_{32,2}h_{23}\right)_{,3}+h_{32}h_{23,2}{\bf(32)}_{,3}
\end{eqnarray}
where
$\textbf{(\emph{ij})}, \textbf{(\emph{ijk})}, \textbf{[\emph{ij}]}\in M_{m_im_i}$ $\left(i, j, k \in \left\{1,2,3\right\}\right)$.

The question of functional relations between covariants is more subtle than for invariants.
\\

{\bf{The general case}:}

Let us consider the following differential operator

\[{\mathbb L}=\left(\begin{array}{cccc}
  I_{m_1}\partial_1 & h_{12}&\ldots & h_{1n}\\
   h_{21} & I_{m_2}\partial_2&\ldots & h_{2n}\\
   \vdots & \vdots & \ddots & \vdots\\
   h_{n1} &h_{n2} &\ldots &I_{m_n}\partial_{n}
  \end{array}
  \right)\]
where the $h_{ij}$ are functions of $x_1$, $x_2$,...,$x_n$ and
the $I_{m_i}$ are unit matrices such that $h_{ij} \in M_{m_im_j}$ and $I_{m_i} \in M_{m_im_i}$
where $i, j \in \left\{1,2, \ldots, n\right\}$.

The gauge transformation
\begin{equation} \mathbb L\mapsto
{\mathbb L}'=g^{-1}\mathbb L g,
\end{equation}
where
\begin{equation}\label{residual}
g=\left(\begin{array}{cccc}
g_1(x_2,x_3,\ldots,x_n) & 0 & \ldots & 0\\
0 & g_2(x_1,x_3,\ldots,x_n ) & & 0\\
\vdots & & \ddots & \vdots \\
0 & \ldots & & g_n(x_1,x_2,\ldots,x_{n-1})
\end{array}\right),
\end{equation}
gives us
\begin{equation}\label{hij}
h_{ij}\mapsto h'_{ij}=g_i^{-1}h_{ij}g_j
\end{equation}
where the $g_i$ are square matrices such that $g_i \in M_{m_im_i}$.

The relations (\ref{hij}) gives us the following matrix covariants:
\begin{eqnarray}
 {\textbf{[\emph{ij}]}}&=&{\textbf{(\emph{ij})}}\left(h_{ij,j}h_{ji}\right)_{,i}+h_{ij}h_{ji,j}{\textbf{(\emph{ij})}}_{,i}\\
 \text{\boldmath {$(i_1i_2i_3 \ldots i_n)$}}&=&h_{i_1i_2}h_{i_2i_3}\ldots h_{i_ni_1}
 \end{eqnarray}
 where $\textbf{[\emph{ij}]}$ $\in M_{m_im_i}$ and {\boldmath{$(i_1i_2i_3 \ldots i_n)$}} $\in M_{m_{i_1}m_{i_1}}$.

\section{Conclusions and comments}

In this paper, we have dealt with general hyperbolic systems ${\mathbb L}z=0$. We have used a suitable diagonal gauge matrix $g$,
chosen so that it kills diagonal terms $h_{ii}$ where $i=1, 2, \ldots , n$.
We have also obtained the complete set of invariants for general hyperbolic systems where rank equals dimension by using the gauge transformation
$\mathbb L\mapsto {\mathbb L}'=g^{-1}\mathbb L g$.
Further, we have shown the completeness of a set of simple invariants (reduced invariants).
We have proved that these invariants form a minimal complete set.

We have also considered hyperbolic systems ${\mathbb L}z=0$ where the entries $h_{ij}$ are matrices. In this case, we are interested in covariants. We have obtained matrix covariants for the differential operator ${\mathbb L}$ under the gauge transformation. Here we have examined the case where rank exceeds dimension. The canonical form of ${\mathbb L}$ still suffices where $h_{ii}=0$ and $h_{ij}$ are rectangular matrices. The reduced covariants have been presented but it has not been shown that their invariant traces form a complete set. For example, in the case when $n=2$, we ask the question: Do the covariants ${\bf(12)}$, ${\bf(21)}$, ${\bf[12]}$ and ${\bf[21]}$ form a complete set? The answer depends on the existence of $g(x_1,x_2)$ so that when
\begin{equation}\label{covts}
 \begin{split}
  {\bf(12)}'=g_1^{-1}{\bf(12)}g_1\\
  {\bf(21)}'=g_2^{-1}{\bf(21)}g_2\\
  {\bf[12]}'=g_1^{-1}{\bf[12]}g_1 \hspace{0.1cm}\\
  {\bf[21]}'=g_2^{-1}{\bf[21]}g_2 \hspace{0.1cm}
 \end{split}
\end{equation}
are given then $g$ must satisfy the relation
\begin{eqnarray*}
 g^{-1}{\mathbb L}g={\mathbb L}'.
\end{eqnarray*}
The square matrices ${\textbf{(\emph{ij})}}$, ${\textbf{(\emph{ij})}}'$, ${\textbf{[\emph{ij}]}}$, ${\textbf{[\emph{ij}]}}'$ are thus similar (\ref{covts}) and so possess as equal invariants the traces, say, of their powers: $I_p=Tr {\textbf{(\emph{ij})}}^p$ etc. But equality of such invariants is not sufficient for gauge equivalence of ${\mathbb L}'$ and ${\mathbb L}$. There are also invariants associated with polynomials in ${\textbf{(\emph{ij})}}$ and ${\textbf{[\emph{ij}]}}$ since $g_i{\textbf{(\emph{ij})}}{\textbf{[\emph{ij}]}}={\textbf{(\emph{ij})}}'{\textbf{[\emph{ij}]}}'g_i$ etc,
namely, traces of such polynomials $\left(\text{cf.}~(\ref{trh})\right)$.

Two questions arise for further study:
\begin{enumerate}
\item What relations on the invariants of these general systems correspond to specialisations of ${\mathbb L}$ such as self-adjointness?
\item Can we establish the existence of a complete, minimal set of trace polynomial invariants for the systems of Section \ref{sec3}?
\end{enumerate}


\subsection*{Acknowledgments}
We would like to thank the School of Mathematics and Statistics, University of Glasgow, for hosting one of us (H.Y.) as a Honorary Research Fellow during Summer 2010 and Summer 2011.

\bibliographystyle{amsplain}
\bibliography{99}

\end{document}